\documentclass[aps]{revtex4}
\usepackage{rotate}
\usepackage{epsfig}
\usepackage{psfrag}

\newcommand \be  {\begin{equation}}
\newcommand \bea {\begin{eqnarray} \nonumber }
\newcommand \ee  {\end{equation}}
\newcommand \eea {\end{eqnarray}}
\newcommand{\subs}[1]{{\mbox{\scriptsize #1}}}
\newcommand \Tr {\mbox{Tr}}
\newcommand \rmi {\mbox{i}}

\begin{document}
\title{Financial Applications of Random Matrix 
Theory: Old Laces and New Pieces
}

\author{Marc Potters$^*$, Jean-Philippe Bouchaud$^{\dagger,*}$
and Laurent Laloux$^*$}
\affiliation{
{$^*$ Science \& Finance, Capital Fund Management, 6 Bd Haussmann,}
{75009 Paris France}\\
{$^\dagger$ Commissariat \`a l'Energie Atomique, Orme des Merisiers}\\
{91191 Gif-sur-Yvette {\sc cedex}, France}}

\begin{abstract}
This contribution to the proceedings of the Cracow meeting on `Applications of 
Random Matrix Theory' summarizes a series of studies, some old and others 
more recent on financial applications of Random Matrix Theory (RMT). We first review 
some early results in that field, with particular emphasis on 
the applications of correlation cleaning to portfolio optimisation, and discuss
the extension of the Mar\v{c}enko-Pastur (MP) distribution to a non trivial `true' 
underlying correlation matrix. We then present new results concerning different problems 
that arise in a financial context: (a) the 
generalisation of the MP result to the case of an empirical correlation matrix (ECM)
constructed using exponential moving averages, for which we give a new elegant 
derivation (b) the specific dynamics of the `market' eigenvalue and its 
associated eigenvector, which defines an
interesting Ornstein-Uhlenbeck process on the unit sphere and 
(c) the problem of the dependence of ECM's on the observation 
frequency of the returns and its interpretation in terms of lagged cross-influences.
\end{abstract}

\maketitle

\section{Portfolio theory: basic results}

Suppose one builds a portfolio of $N$ assets with weight $w_i$ on the $i$th asset. 
The (daily) variance of the portfolio return is
given by:
\be
R^2=\sum_{ij} w_i \sigma_i C_{ij} \sigma_j w_j,
\ee
where $\sigma_i^2$ is the (daily) variance of asset $i$ and $C_{ij}$ is the correlation matrix.
If one has predicted gains $g_i$, then the expected gain of the portfolio is $G=\sum w_i g_i$.

In order to measure and optimize the risk of this portfolio, one therefore has to come up with a 
reliable estimate of the correlation matrix $C_{ij}$. This is difficult in general \cite{prl,book} 
since one has to 
determine of the order of $N^2/2$ coefficients out of $N$ time series of length $T$, and in general 
$T$ is not much larger than $N$ (for example, 4 years of data give 1000 daily returns, and the typical
size of a portfolio is several hundred stocks.) We will denote, in the following, $q=N/T$; an accurate
determination of the true correlation matrix will require $q \ll 1$.
If $r^i_t$ is the daily return of stock $i$ at time $t$, 
the empirical variance of each stock is: 
\be
\sigma^2_i=\frac{1}{T}\sum_{t}^T \left(r^i_t\right)^2,
\ee
and can be assumed for simplicity to be perfectly known (its relative mean square-error is $(2+\kappa)/T$, where 
$\kappa$ is the kurtosis of the stock, known to be typically 
$\kappa \approx 3$). In the above definition, we have, as usual, neglected the daily mean return, small compared to daily fluctuations. 
The empirical correlation matrix is obtained as:
\be\label{defE}
E_{ij}=\frac{1}{T}\sum_{t}^T x_t^i x_t^j; \qquad x^i_t\equiv r^i_t/\sigma_i.
\ee
If $T<N$, $\mathbf{E}$ has rank $T<N$, and has $N-T$ zero eigenvalues. 
Assume there is a ``true'' correlation matrix $\mathbf{C}$ from 
which past and future $x^i_t$ 
are drawn. The risk of a portfolio constructed {\it independently} of the past
realized $x^i_t$ is faithfully 
measured by:
\be
\left\langle R^2_E\right\rangle=\frac{1}{T}\sum_{ijt}w_i\sigma_i\left\langle x^i_t x^j_t\right\rangle
w_j\sigma_j \approx \sum_{ij} w_i \sigma_i C_{ij} \sigma_j w_j.
\ee
Because the portfolio is not constructed using $\mathbf{E}$, this estimate is unbiased and the relative 
mean square-error is small ($\sim 1/T$). Otherwise, the $w$'s would depend on the observed $x$'s and, as
we show now, the result can be very different.

Problems indeed arise when one wants to estimate the risk of an optimized portfolio, resulting from a 
Markowitz optimization scheme, which gives the portfolio with maximum expected return for a given risk or equivalently,
the minimum risk for a given return ($G$):
\be
w_i\sigma_i = G\frac{\sum_jC^{-1}_{ij} g_j/\sigma_j}{\sum_{ij}g_i/\sigma_i
C^{-1}_{ij} g_j/\sigma_j}
\ee
From now on, we drop $\sigma_i$ (which can always be absorbed in $g_i$ and $w_i$).
In matrix notation, one has:
\be
\mathbf{w}_C = G\frac{\mathbf{C}^{-1}\mathbf{g}}{\mathbf{g}^{T}\mathbf{C}^{-1}\mathbf{g}}
\ee
The question is to estimate the risk of this optimized portfolio, and in particular to understand the 
biases of different possible estimates. We define the following three quantities:
\begin{itemize} 
\item The ``In-sample'' risk, corresponding to the risk of the optimal portfolio over the period used
to construct it:
\be
R^2_\subs{in}=\mathbf{w}_E^{T}\mathbf{E}\mathbf{w}_E = \frac{G^2}{\mathbf{g}^{T}\mathbf{E}^{-1}\mathbf{g}}
\ee
\item The ``true'' minimal risk, which is the risk of the optimized portfolio in the ideal world where $\mathbf{C}$
would be perfectly known:
\be
R^2_\subs{true}=\mathbf{w}_C^{T}\mathbf{C}\mathbf{w}_C =\frac{G^2}{\mathbf{g}^{T}\mathbf{C}^{-1}\mathbf{g}}
\ee
\item  The ``Out-of-sample'' risk which is the risk of the portfolio constructed using $\mathbf{E}$, 
but observed on the next period of time: 
\be
R^2_\subs{out}=\mathbf{w}_E^{T}\mathbf{C}\mathbf{w}_E =
G^2 \frac{\mathbf{g}^{T}\mathbf{E}^{-1}\mathbf{CE}^{-1}\mathbf{g}}
{(\mathbf{g}^{T}\mathbf{E}^{-1}\mathbf{g})^2}
\ee
\end{itemize}
From the remark above, the result is expected to be the same (on average) 
computed with $\mathbf{C}$ or with $\mathbf{E}'$, the
ECM corresponding to the second time period. Since $\mathbf{E}$ is a noisy estimator of $\mathbf{C}$ 
such that $\langle\mathbf{E}\rangle=\mathbf{C}$, one can use a convexity argument for the inverse of 
positive definite matrices to show that in general:
\be
\langle\mathbf{g}^{T}\mathbf{E}^{-1}\mathbf{g}\rangle \geq \mathbf{g}^{T}\mathbf{C}^{-1}\mathbf{g}
\ee
Hence for large matrices, for which the result is self-averaging:
\be
R^2_\subs{in} \leq R^2_\subs{true}.
\ee
By optimality, one clearly has that:
\be
R^2_\subs{true} \leq R^2_\subs{out}.
\ee
These results show that the out-of-sample risk of an optimized portfolio is larger (and in practice, much larger,
see Fig 1) than the in-sample risk, which itself is an underestimate of the true minimal risk. This is a 
general situation: using past returns to optimize a strategy always leads to over-optimistic results because 
the optimization adapts to the particular realization of the noise, and is unstable in time. In the case where  
the true correlation matrix is $\mathbf{C}=\mathbf{1}$, one can show that
\cite{Kondor}:
\be
R^2_\subs{true} =\frac{G^2}{\mathbf{g}^{T}\mathbf{g}}\qquad
\mbox{{ and}}\qquad
R^2_\subs{in}={R^2_\subs{true}}\sqrt{1-q}=R^2_\subs{out}{(1-q)}
\ee
Only in the limit $q \to 0$ will these quantities coincide, which is expected since in this case the measurement
noise disappears. In the other limit $q \to 1$, the in-sample risk becomes zero since it becomes possible to find 
eigenvectors (portfolios) with zero eigenvalues (risk), simply due to the lack of data.

\begin{figure}
\begin{center}
\psfig{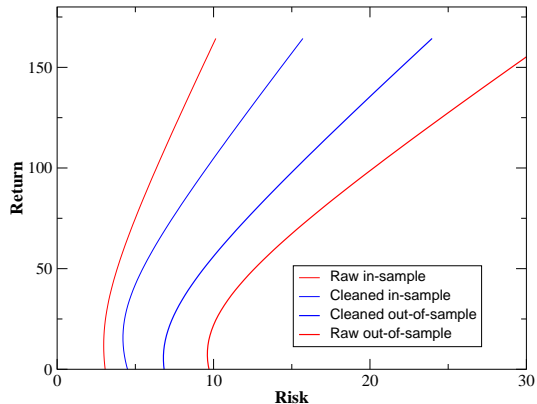} 
\end{center}
\caption{In sample (left curves) and out of sample (right curves) 
portfolio risk along the optimal 
curve in the risk return plane. Red lines are with the raw empirical matrix, blue lines 
with the cleaned matrix using RMT, showing how the risk underestimation can be
reduced by matrix cleaning. From \cite{dublin}.
}
\label{Fig1}
\end{figure}

\section{Matrix cleaning and RMT}

How should one `clean' the empirical correlation matrix to avoid, as much as possible, such biases in the estimation
of future risk? In order to get some general intuition on this problem, 
let us rewrite the Markowitz solution 
in terms of the eigenvalues $\lambda_k$ and eigenvectors $V_i^k$ of the 
correlation matrix:
\be
w_i \propto \sum_{kj} \lambda_k^{-1} V_i^k V_j^k g_j \equiv 
g_i + \sum_{kj} (\lambda_k^{-1}-1) V_i^k V_j^k g_j
\ee
The first term corresponds to the naive solution: one should invest proportionally to the expected gain 
(in units where $\sigma_i=1$). The correction term means that the weights 
of eigenvectors with $\lambda > 1$ are suppressed, 
whereas the weights of eigenvectors with $\lambda < 1$ are enhanced. Potentially, the optimal Markowitz solution 
allocates a very large weight to small eigenvalues, which may be entirely dominated by measurement noise and
hence unstable. A very naive way to avoid this is to go for the naive solution $w_i \propto g_i$, but with the 
$k^*$ largest eigenvectors projected out:
\be
w_i \propto  
g_i - \sum_{k\leq k^*;j} V_i^k V_j^k g_j,
\ee
so that the portfolio is market neutral (the largest eigenvector correspond to
a collective market mode, $V_i^1 \approx 1/\sqrt{N}$) 
and sector neutral (other large eigenvectors 
contain sector moves). Since most of the volatility is contained in the 
market and sector modes, the above portfolio is already quite good in terms of risk. More elaborated ways aim at retaining all 
eigenvalues and eigenvectors, but after having somehow cleaned them. A well known cleaning corresponds to the so-called
``shrinkage estimator'': the empirical matrix is shifted `closer' to the identity matrix. This is
a Bayesian method that assumes that the prior empirical matrix is the identity, again only justified if
the market mode has been subtracted out. More explicitely:
\be
\mathbf{E}_c = \alpha \mathbf{E} +(1-\alpha) \mathbf{1} \qquad\mbox{{so}}\qquad
\lambda^k_c = 1 + \alpha (\lambda^k-1),
\ee
where the subscript $c$ corresponds to the cleaned objects. This method involves the parameter $\alpha$
which is undetermined, but somehow related to the expected signal to noise ratio. If the signal is 
large, $\alpha\to 1$, and $\alpha \to 0$ if the noise is large. Another possible interpretation is through a
constraint on the effective number of assets in the portfolio, defined as $(\sum_i w_i^2)^{-1}$ \cite{book}. Constraining this
number to be large (ie imposing a minimal level of diversification) is equivalent to choosing $\alpha$ small. 

Another route is eigenvalue cleaning, first suggested in \cite{prl,dublin}, where one replaces all low lying eigenvalues
with a unique value, and keeps all the high eigenvalues corresponding to meaningful economical information (sectors):
\be
\lambda^k_c = 1-\delta \qquad\mbox{{if}}\qquad k>k^*,
\qquad
\lambda^k_c = \lambda^k_E  \qquad\mbox{{if}}\qquad k\leq k^*,
\ee
where $k^*$ is the number of meaningful number of sectors and $\delta$
is chosen such that the trace of the correlation matrix is exactly
preserved. How should $k^*$ be chosen? The idea developed in 
\cite{prl,dublin} is to use Random Matrix Theory to determine the theoretical edge of the `random' part of the
eigenvalue distribution, and to fix $k^*$ such that $\lambda^{k^*}_E$ is close to this edge.

What is then the spectrum of a random correlation matrix? The answer is known in several cases, due to the work of Mar\v{c}enko and Pastur \cite{marcenkopastur} and others 
\cite{baisilverstein,sengupta,burdagorlich}. We briefly recall the results and some
elegant methods to derive them, with special emphasis on the problem of the largest eigenvalue, 
which 
we expand on below. Consider an empirical correlation matrix $\mathbf{E}$ of $N$ assets using $T$ data points, 
both very large, with $q=N/T$ finite. Suppose that the true correlations are given by $\langle x^i_t x^j_{t'} \rangle = C_{ij}\delta_{tt'}$. This defines the Wishart ensemble \cite{Wishart}. 
In order to find the eigenvalue density, one introduces the resolvent:
\be
G(z)=\frac{1}{N}\Tr\left[(z\mathbf{I}-\mathbf{E})^{-1}\right]
\ee
from which one gets:
\be\label{im_part}
\rho(\lambda) = \lim_{\epsilon \to 0} \frac{1}{\pi} \Im\left( G(\lambda-\rmi\epsilon)
\right).
\ee
The simplest case is when $\mathbf{C}=\mathbf{I}$. Then, $\mathbf{E}$ is a sum 
of rotationally invariant 
matrices $\delta E^t_{ij}=(x^i_t x^j_t)/T$. The trick in that case is to 
consider the so-called Blue function,
inverse of $G$, i.e. such that $B(G(z))=z$. The quantity $R(x)=B(x)-1/x$ 
is the `R-transform' of $G$, and is known to be
additive \cite{marcenkopastur,voiculescu,zee}. Since 
$\delta \mathbf{E}^t$ has one eigenvalue 
equal to $q$ and $N-1$ zero eigenvalues, one has:
\be
\delta G_t(z)=\frac{1}{N}\left(\frac{1}{z-q}+\frac{N-1}{z}\right)
\ee
Inverting $\delta G_t(z)$ to first order in $1/N$, 
\be
\delta B_t(x)=\frac{1}{x}+\frac{q}{N(1-qx)}\quad\longrightarrow \quad B_E(x)=\frac{1}{x}+\frac{1}{(1-qx)},
\ee
and finally
\be
G_E(z)=\frac{(z+q-1)-\sqrt{(z+q-1)^2-4zq}}{2 z q},
\ee
which finally reproduces the well known Mar\v{c}enko-Pastur (MP) result:
\be
\rho(\lambda) = \frac{\sqrt{4\lambda q - (\lambda+q-1)^2}}{2\pi \lambda q}
\ee

The case of a general $\mathbf{C}$ cannot be directly written as a sum of ``Blue'' functions, but
one can get a solution using the Replica trick or by summing planar diagrams, which gives the 
following relation between resolvents: \cite{baisilverstein,sengupta,burdagorlich}
\be\label{burda}
zG_E(z) = Z G_C(Z) \qquad\mbox{{where}}\qquad Z = \frac{z}{1+q(z G_E(z) -1)},
\ee 
from which one can easily obtain $\rho(\lambda)$ numerically 
for any choice of $\mathbf{C}$ \cite{burdagorlich}. [In fact, this result can even be obtained from
the original Mar\v{c}enko-Pastur paper by permuting the role of the appropriate matrices]. 
The above result does however {\it not} apply when $\mathbf{C}$ has isolated eigenvalues, and
only describes continuous parts of the spectrum. For example, if one considers a matrix $\mathbf{C}$
with one large eigenvalue that is separated from the `Mar\v{c}enko-Pastur sea', 
the statistics of this isolated
eigenvalue has recently been shown to be Gaussian \cite{BenArous} (see also below), with a width $\sim T^{-1/2}$, much smaller than 
the uncertainty on the bulk eigenvalues ($\sim q^{1/2}$). A naive application of Eq. (\ref{burda}),
on the other hand, would give birth to a `mini-MP' distribution around the top eigenvalue. This
would be the exact result only if the top eigenvalue of $C$ had a degeneracy proportional to $N$. 

From the point of view of matrix cleaning, however, these results show that: 
(i) the expected edge of 
the bulk, that determines $k^*$, obviously depends on the prior one has for $\mathbf{C}$. 
The simplest case where $\mathbf{C}=\mathbf{I}$ was investigated in particular in 
\cite{prl,Plerou}, 
with the results shown in Fig 2. 
Other choices are however possible and could lead to an improved cleaning algorithm; (ii) the uncertainty
on large eigenvalues is much less than that on the bulk eigenvalues, meaning that 
the bulk needs a lot of shrinkage, but the bigger eigenvalues less so -- at variance with the naive
shrinkage procedure explained above. An alternative route may consist in using the `power mapping' method
proposed by Guhr \cite{Guhr} or clustering methods \cite{Lillo}.

\begin{figure}
\begin{center}
\psfig{file=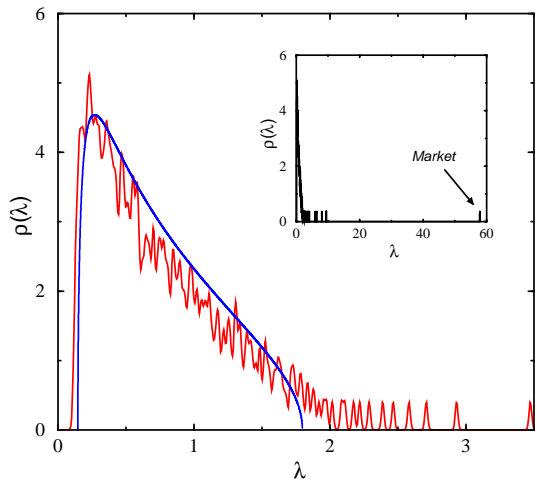,width=7cm} 
\end{center}
\caption{Empirical eigenvalue density for 406 stocks from the S\&P 500, and fit using the
MP distribution. Note (i) the presence of one very large eigenvalue, corresponding to
the market mode (see section IV) and (ii) the MP fit reveals systematic deviations, 
suggesting a non trivial structure of the true correlation matrix, even after sector modes
have been accounted for (see \cite{burdagorlich,us}).
}
\label{Fig2}
\end{figure}

\section{EWMA Empirical Correlation Matrices}

Consider now the case where $\mathbf{C}=\mathbf{I}$  but where the Empirical matrix is computed using an exponentially 
weighted moving average (EWMA). More precisely:
\be
E_{ij}=\epsilon\sum_{t'=-\infty}^{t-1}(1-\epsilon)^{t-t'} x_i^{t'} x_j^{t'} 
\ee
with $\epsilon=1/T$.
Such an estimate is standard practice in finance. Now, as an ensemble $E_{ij}$ satisfies $E_{ij}=(1-\epsilon) E_{ij} + 
\epsilon x_i^{t} x_j^{t}$. We again invert the resolvent of $\delta \mathbf{E_t}$ 
to find the elementary R-function,
\be
\delta B_t(x)=\frac{1}{x}+R_t(x)\qquad\mbox{{with}}\qquad R_t(x)=\frac{q}{N(1-qx)}
\ee
Using the scaling properties of $G(z)$ we find for $R(x)$:
\be
R_{aC}(x)=aR_C(ax).
\ee
This allows one to write:
\be
R_E(x)=R_{(1-\epsilon) E}(x) + R_t(x) = (1-q/N) R_E((1-q/N)x)+\frac{q}{N(1-qx)}
\ee
To first order in $1/N$, one now gets:
\be
R(x)+xR'(x)+\frac{q}{1-qx}=0 \longrightarrow R(x)=-\frac{\log(1-qx)}{qx}.
\ee
Going back to the resolvent to find the density, we finally get the result first obtained 
in \cite{PKP}:
\be
\rho(\lambda) = \frac{1}{\pi} \Im G(\lambda)\quad\mbox{{ where $G(\lambda)$ solves }}
\quad\lambda q G=q-\log(1-qG)
\ee
This solution is compared to the standard MP distribution in Fig 3.

Another nice properties of Blue functions is that they can be used to find the edges
of the eigenvalue spectrum ($\lambda_\pm$). One has:\cite{zee}
\be
\lambda_\pm =B(x_\pm)\qquad\mbox{where}\qquad B'(x_\pm)=0
\ee
In the case at hand, by evaluating $B(x)$ when $B'(x)=0$ we can write 
directly an equation whose solutions are the spectrum edges ($\lambda_\pm$)
\be
\lambda_\pm=\log(\lambda_\pm)+q+1
\ee
When $q$ is zero, the spectrum is a $\delta$ in $1$ as expected. But as the noise increases (or the characteristic
time decreases) the lower edge approach zero very quickly as $\lambda_-\sim\exp(-q)$. Although there are no exact zero eigenvalues for EWMA matrices, 
the smallest eigenvalue is very close to zero.  

\begin{figure}
\begin{center}
\psfig{file=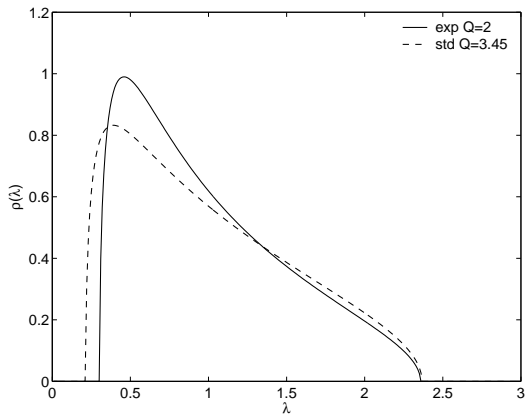,width=7cm} 
\end{center}
\caption{Spectrum of the exponentially weighted
random matrix with $q\equiv N\epsilon=1/2$ and the spectrum of the standard
Wishart random matrix with $q\equiv N/T=1/3.45$, chosen to have the same upper edge. 
From \cite{PKP}.
}
\label{Fig3}
\end{figure}

\section{Dynamics of the top eigenvalue and eigenvector}

As mentioned above, it is well known that financial covariance matrices
are such that the largest eigenvalue is 
well separated from the `bulk', where all other eigenvalues reside. 
The financial interpretation of this large 
eigenvalue is the so-called `market mode': in a first approximation, all stocks move together, up or down. One can state 
this more precisely in the context of the one factor model, where the $i$th stock return at time $t$ is written
as:
\be
r_t^i = \beta_i \phi_t + \varepsilon_t^i,
\ee
where the market mode $\phi_t$ is common to all stocks through their market exposure $\beta^i$ and the  
$\varepsilon_t^i$ are idiosyncratic noises, uncorrelated from stock to stock. Within such a model, the covariance 
matrix reads:
\be
C_{ij} = \beta_i \beta_j \sigma_\phi^2 + \sigma_i^2 \delta_{ij}.
\ee
When all $\sigma_i$'s are equal, this matrix is easily diagonalized; for $N$ stocks, its largest eigenvalue is
$\Lambda_0= (\sum_j \beta_j^2) \sigma_\phi^2 + \sigma^2$ and is of order $N$, and all the other $N-1$ eigenvalues
$\Lambda_\alpha$ are equal to $\sigma^2$. The largest eigenvalue corresponds to the eigenvector $\beta_i$. More 
generally, the largest eigenvalue $\Lambda_0$, normalized by the average square volatility of the stocks, can be seen as a 
proxy for the average interstock correlation.

A natural question, of great importance for portfolio management, or 
dispersion trading (option strategies based on the implied average 
correlation between stocks), 
is whether $\Lambda_0$ and the $\beta$'s are stable in time. Of course, 
the largest eigenvalue and eigenvector of the
empirical correlation matrix will be, as discussed at length above, affected by measurement noise. Can one make 
predictions about the fluctuations of both the largest eigenvalue and the corresponding eigenvector induced by 
measurement noise? This would help separating a true evolution in time of the average stock correlation 
and of the market exposure of each stock from one simply related to measurement noise. We shall see that such a 
decomposition seems indeed possible in the limit where $\Lambda_0 \gg \Lambda_\alpha$. 

We will assume, as in the previous section, that the covariance matrix 
is measured through an exponential moving average of the returns. 
This means that
the matrix ${\mathbf E}$ evolves in time as:
\be
E_{ij,t} = (1 - \epsilon) E_{ij,t-1} + \epsilon r_t^i r_t^j.
\ee
The {\it true} covariance matrix $C_{ij} = \langle r^i r^j \rangle$ 
is assumed to be time independent -- this will give
us our benchmark hypothesis -- with its largest eigenvalue $\Lambda_0$ associated to the normalized eigenvector $\Psi_0$. In this section we 
deal with covariance
matrices instead of correlation matrices for simplicity, but most results 
should carry over to this latter case as well.

Denoting as $\lambda_{0t}$ the largest eigenvalue of $E_t$ associated to $\psi_{0t}$, standard perturbation theory, valid
for $\epsilon \ll 1$, gives:
\be
\lambda_{0t} = (1 - \epsilon) \lambda_{0t-1} + \epsilon \langle \psi_{0t-1} | C | \psi_{0t-1} \rangle 
+ \epsilon \langle \psi_{0t-1} | \eta_{t} | \psi_{0t-1} \rangle,
\ee
with $\eta_{ij} = r^i r^j - \langle r^i r^j \rangle$. We will suppose for simplicity that the returns are Gaussian, yielding:
\be \label{etavar}
\langle \eta_{ij}\eta_{k\ell} \rangle = C_{ik}C_{j\ell} + C_{i\ell}C_{jk}.
\ee
In the limit where $\Lambda_0$ becomes much larger than all 
other eigenvectors, the above equation simplifies to:
\be\label{OU}
\lambda_{0t} \approx (1 - \epsilon) \lambda_{0t-1} 
+ \epsilon \cos^2 \theta_{t-1} \Lambda_0 \left[ 1 + \xi_t \right],  
\ee
where $\cos \theta_t \equiv \langle \psi_{0t} | \Psi_0 \rangle$ and $\xi_t$ is a random noise term of mean zero and 
variance equal to $2$. This noise becomes Gaussian in the limit of large matrices, leading to a Langevin 
equation for $\lambda_0$:
\be\label{Orn1}
\frac{d\lambda_0}{dt} = \epsilon(\cos^2 \theta \Lambda_0- \lambda_{0}) + 
\epsilon \cos^2 \theta \xi_t.
\ee
We have neglected in the above equation a deterministic term equal to 
$\epsilon \sin^2 \theta \Lambda_1$, which will turn out to be a factor
$(\Lambda_1/\Lambda_0)^2$ smaller than the terms retained in Eq. (\ref{OU}).

We now need to work out an equation for the projection of the instantaneous eigenvector 
$\psi_{0t}$ on the true eigenvector $\Psi_0$. This can again be done using perturbation theory, which gives, in 
braket notation:
\bea\nonumber
| \psi_{0t} \rangle &=& | \psi_{0t-1} \rangle + \epsilon \sum_{\alpha \neq 0} 
\frac{\langle \psi_{\alpha t-1} | r_t r_t| \psi_{0t-1} \rangle}{\lambda_{0t-1} - \lambda_{\alpha t-1}}
| \psi_{\alpha t-1} \rangle\\ 
&\approx& | \psi_{0t-1} \rangle + \epsilon \frac{r_t r_t| \psi_{0 t-1} \rangle}{\lambda_{0t-1}}
- \epsilon \frac{\langle \psi_{0t-1} | r_t r_t| \psi_{0t-1} \rangle}{\lambda_{0t-1}}
| \psi_{0t-1} \rangle,
\eea
where we have used the fact that the basis of eigenvectors is complete. It is clear by inspection that the 
correction term is orthogonal to $| \psi_{0t-1} \rangle$, so that $| \psi_{0t} \rangle$ is still, to first 
order in $\epsilon$, normalized. Let us now decompose the matrix $r_t r_t$ into its average part $C$ and
the fluctuations $\eta$, and first focus on the former contribution. Projecting the above equation on $< \Psi_0 |$ leads
to the deterministic part of the evolution equation for $\cos \theta_t$:
\be
\cos \theta_t \approx \cos \theta_{t-1} + \epsilon \cos \theta_{t-1} \frac{\Lambda_0}{\lambda_{0t-1}}
- \epsilon \cos^3 \theta_{t-1} \frac{\Lambda_0}{\lambda_{0t-1}},
\ee
where we have neglected the contribution of the small 
eigenvalues compared to $\Lambda_0$, which is again a factor 
$(\Lambda_1/\Lambda_0)^2$ smaller. In the continuous 
time limit $\epsilon \to 0$, this equation can be rewritten as:
\be\label{Orn}
\frac{d\theta}{dt} = - \frac{\epsilon \Lambda_0}{2\lambda_{0t}}  \sin 2\theta,
\ee
and describes a convergence of the angle $\theta$ towards $0$ or $\pi$ -- clearly, $\Psi_0$ and $-\Psi_0$ are 
equivalent. It is the noise term $\eta$ that will randomly push the instantaneous eigenvector away from its 
ideal position, and gives to $\theta$ a non-trivial probability distribution. Our task is therefore to compute the
statistics of the noise, which again becomes Gaussian for large matrices, so that we only need to compute its 
variance. Writing $| \psi_{0t} \rangle = \cos \theta_t | \Psi_{0} \rangle + \sin \theta_t | \Psi_{1t} \rangle$, 
where $| \Psi_{1t} \rangle$ is in the degenerate eigenspace corresponding to small eigenvalues $\Lambda_1$, 
and using Eq. (\ref{etavar}), we find that the noise term $\zeta_t$ to be added to Eq. (\ref{Orn}) is given by:
\be
\langle \zeta_t^2 \rangle 
\approx \frac{\epsilon^2}{\lambda_{0t}^2} \left[2 \Lambda_0^2 
\cos^2 \theta_t \sin^2 \theta_t + \Lambda_0 \Lambda_1 \cos^2 2\theta_t \right],
\ee
where we have kept the second term because it becomes the dominant source of noise when $\theta \to 0$, but neglected a term in $\Lambda_1^2$. The 
eigenvector $\psi_0$ therefore undergoes an Ornstein-Uhlenbeck like motion 
on the unit sphere.
One should also note that the two sources of noise $\xi_t$ and $\zeta_t$ are not independent. Rather, one has, neglecting $\Lambda_1^2$ terms:
\be\label{correl}
2\langle \xi_t \zeta_t \rangle \approx \Lambda_0 \cos^2 \theta_t \sin 2\theta_t - \Lambda_1 \sin 4\theta_t  
\ee
In the continuous time limit, we therefore have two coupled Langevin equations for the top eigenvalue $\lambda_{0t}$
and the deflection angle $\theta_t$. In the limit $\Lambda_1 \ll \Lambda_0$, the stationary solution for the angle can be 
computed to be:
\be
P(\theta) = {\cal N} \left[\frac{1 + \cos 2\theta (1 - \frac{\Lambda_1}{\Lambda_0})}
{1 - \cos 2\theta (1 - \frac{\Lambda_1}{\Lambda_0})}\right]^{1/4\epsilon}
\ee
As expected, this distribution is invariant when $\theta \to \pi - \theta$, since $-\Psi_0$ is also a top eigenvector.
In the limit $\Lambda_1 \ll \Lambda_0$, one sees that the distribution becomes peaked around $\theta=0$ and $\pi$.
For small $\theta$, the distribution becomes Gaussian:
\be
P(\theta) \approx \frac{1}{\sqrt{2 \pi \epsilon \frac{\Lambda_1}{\Lambda_0}}}
 \exp\left(-\frac{\theta^2}{2 \epsilon 
\frac{\Lambda_1}{\Lambda_0}}\right),
\ee
leading to $\langle \cos^2 \theta \rangle 
\approx 1 - \epsilon {\Lambda_1}/2{\Lambda_0}$ The angle $\theta$ is less 
and less fluctuating as $\epsilon \to 0$ (as expected) but also as $\Lambda_1/\Lambda_0 \to 0$: a large separation of
eigenvalues leads to a well determined top eigenvector. In this limit, the
distribution of $\lambda_0$ also becomes Gaussian (as expected from general results 
\cite{BenArous}) and one finds, to leading order:
\be
\langle \lambda_0 \rangle \approx \Lambda_0 - \epsilon {\Lambda_1}/2; \qquad 
\langle(\delta \lambda_0)^2\rangle \approx \epsilon.
\ee
Therefore, we have shown that in the limit of large averaging time and one large top eigenvalue (a situation 
approximately realized for financial markets), the deviation from the true top eigenvalue 
$\delta \lambda_0$ and 
the deviation angle $\theta$ are independent 
Gaussian variables (the correlation between them indeed becomes 
zero as can be seen using Eq. (\ref{correl}) in that limit, both following Ornstein-Uhlenbeck
processes.  

\begin{figure}
\begin{center}
\psfig{file=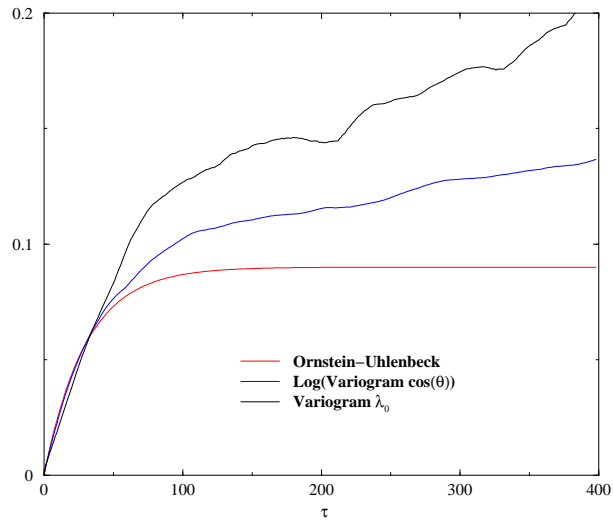,width=7cm,angle=270} 
\end{center}
\caption{Variogram of the top eigenvalue of the {\it correlation} matrix 
$\lambda_0$ (top curve) and (log)-variogram 
of the top eigenvector, $\ln\left\langle \langle \psi_{0t+\tau}|\psi_{0t}
\rangle \right \rangle$, as a function of the time lag $\tau$, for US stocks,
and for an exponential moving average parameter $\epsilon=1/30$ days. 
We show for comparison (bottom curve) the Ornstein-Uhlenbeck behaviour expected for both 
quantities in the case of a time independent underlying correlation structure.  
}
\label{Fig4}
\end{figure}

From these results one directly obtains the variogram of the top eigenvalue as:
\be
\langle[\lambda_{t+\tau}-\lambda_t]^2\rangle
=2 \epsilon \left(1 - \exp(-\epsilon \tau)\right).
\ee
This is the result expected in the absence of 
a `true' dynamical evolution of the structure
of the matrix. From Fig. 4, one sees that there is clearly an additional
contribution, hinting at a real evolution of the strength of the 
average correlation with time. 
One can also work out the average overlap of the top eigenvector
with itself as a function of time lag, $E(\langle \psi_{0t+\tau}|\psi_{0t}
\rangle)$. Writing again $| \psi_{0t} \rangle = 
\cos \theta_t | \Psi_{0} \rangle + \sin \theta_t | \Psi_{1t} \rangle$, one sees that an equation for 
the evolution of the transverse component $| \Psi_{1t} \rangle$ is a priori
also needed. It is easy to show, following the same method as above, that
the evolution of the angle $\phi_t$ made by the component with a fixed
direction is a free diffusion with a noise term of order $\epsilon 
\Lambda_1/\Lambda_0$. Therefore, on the time scale $\epsilon^{-1}$ over
which $\theta_t$ evolves,  $| \Psi_{1t} \rangle$ can be considered to 
be quasi-constant, leading to:
\be 
\left\langle \langle \psi_{0t+\tau}|\psi_{0t}
\rangle \right \rangle \approx E(\cos(\theta_t - \theta_{t+\tau})) \approx 
1 - \epsilon \frac{\Lambda_1}{\Lambda_0} (1 - \exp(-\epsilon \tau)).
\ee
Any significant deviation from the above law would, again, indicate a true 
evolution of the market structure. Again, Fig. 4, provides some evidence of
such an evolution, although weaker than that of the top eigenvalue $\lambda_0$.

\section{Frequency dependent correlation matrices}

The very definition of the correlation matrix {\it a priori} depends on the time scale over which returns are 
measured. The return on time $\tau$ for stock $i$ is defined as: $r_{i,\tau}(t) = \ln p_i(t+\tau) - \ln p_i(t)$,
where $p_i(t)$ is the price at time $t$. The correlation matrix is then defined as:
\be
C_{ij}(\tau)=\frac{\langle r_{i,\tau}(t) r_{j,\tau}(t)\rangle_c}{\sigma_i \sigma_j}
\ee
A relatively well known effect is that the average inter-stock correlation grows with the observation time scale --
this is the so-called Epps effect \cite{Epps,Reno}. For example, for a collection of stocks from the FTSE index, one finds, in the period 1994-2003:
\be
\langle C_{i \neq j}(5') \rangle \approx 0.06; \qquad  
\langle C_{i \neq j}(1h) \rangle \approx 0.19; \qquad
\langle C_{i \neq j}(1d) \rangle \approx 0.29
\ee
Besides the change of the average correlation level, there is also a change of structure of the correlation matrix: the full eigenvalue distribution distribution 
changes with $\tau$. A trivial effect is that by increasing the observation frequency one also increases the number of observations; the parameter $q$ defined
above decreases and the noise band is expected to shrink. This, at first sight, appears to be a nice way to
get rid of the observation noise in the correlation matrix (see \cite{Iori} for a related discussion). Unfortunately, the problem (or the interesting effect,
depending on the standpoint) is that this is accompanied by a true modification of the correlations, for which we will provide a 
model below. In particular one observes the 
emergence of a larger number of eigenvalues leaking out from the bulk of the 
eigenvalue spectrum (and corresponding to `sectors') as the 
time scale $\tau$ increases. This effect was also noted by Mantegna \cite{Mantegna}: the structure of the minimal spanning tree
constructed from the correlation matrix evolves from a `star like' structure for small $\tau$'s (several minutes),
meaning that only the market mode is present, to a fully diversified tree at the scale of a day. Pictorially, the
market appears as an embryo which progressively forms and differentiates with time. 

The aim of this section is to introduce a simple model of lagged cross-influences that allows one 
to rationalize the mechanism leading to such an evolution of the correlation matrix. Suppose that the 
return of stock $i$ at time $t$ is influenced in a causal way by return of stock $j$ at all previous times $t' < t$. 
The most general linear model for this reads:
\be
r_{i,1}(t) = \xi_i(t) + 
\sum_j \int_{-\infty}^t {\rm d}t' K_{ij}(t-t') r_{j,1}(t')
\qquad \langle  \xi_i(t) \xi_j(t') \rangle = D_i \delta_{ij} \delta(t-t')
\ee
Here $\tau=1$ is the shortest time scale -- say a few seconds. The kernel $K_{ij}$ is in general non-symmetric and
describes how the return of stock $j$ affects, on average, that of stock $i$ a certain time later. 
We will define the lagged correlation ${\cal C}_{ij}(t-t')$ by:
\be
{\cal C}_{ij}(t-t') = \langle r_{i,1}(t) r_{j,1}(t') \rangle.
\ee
This matrix is, for $t \neq t'$, not symmetric; however, one has obviously ${\cal C}_{ij}(t-t')=
{\cal C}_{ji}(t'-t)$. These lagged correlations were already studied in \cite{Kertesz}.
Going to Fourier space, one finds the Fourier transform of the covariance matrix ${\cal C}_{ij}(\omega)=
{\cal C}_{ji}(-\omega)$:
\be
{\cal C}_{ij}(\omega) = \sum_k (1 - K(\omega))^{-1}_{ik}
(1 - K(-\omega))^{-1}_{jk} D_k 
\ee 
where $K(\omega)$ is the Fourier transform of $K(\tau)$ with by convention $K(\tau < 0)=0$. 
When cross-correlations are small, which is justified provided the $r_{i,1}(t)$ corresponds
to residual returns, where the market has been subtracted, the relation between ${\cal C}_{ij}$ and $K_{ij}$ 
becomes quite simple and reads, for $\tau > 0$:
\be
{\cal C}_{ij}(\tau) = D_j K_{ij}(\tau).
\ee
This equation allows one, in principle, to determine $K_{ij}(\tau)$ from the 
empirical observation of the lagged correlation matrix. Suppose for simplicity that the influence kernel takes the form 
$K_{ij}(\tau)=K_{ij}^0 e^{-\Gamma_{ij} \tau}$, then $K_{ij}(\omega)=K_{ij}^0/(i\omega + \Gamma_{ij})$.
In this model, the primary object is the influence matrix $K$ which has a much richer structure than the 
correlation matrix: each element defines an {\it influence strength} $K^0$ and an {\it synchronisation time} $\Gamma^{-1}$.
In fact, as shown in Figs. 5 and 6, fitting the empirical data requires that $K_{ij}$ is parameterized by a sum of 
at least {\it two} exponentials, one corresponding to a time scale of minutes, and a second one of hours; 
sometimes the influence strength corresponding to these two time scales have opposite signs. Pooling together
the parameters corresponding to different pairs of stocks, we find, as might have been expected, that 
strongly coupled stocks (large $K^0$) have short synchronisation times $\Gamma^{-1}$.

\begin{figure}
\begin{center}
\psfig{file=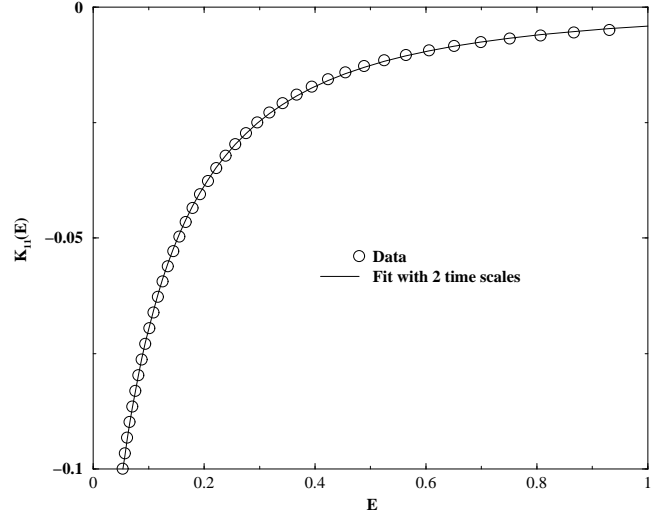,width=7cm,angle=270} 
\end{center}
\caption{Typical self-influence kernel $K_{ii}(E)$ in Laplace space and 
fit with the Laplace transforms of the sum of two exponentials.
}
\label{Fig5}
\end{figure}


\begin{figure}
\begin{center}
\psfig{file=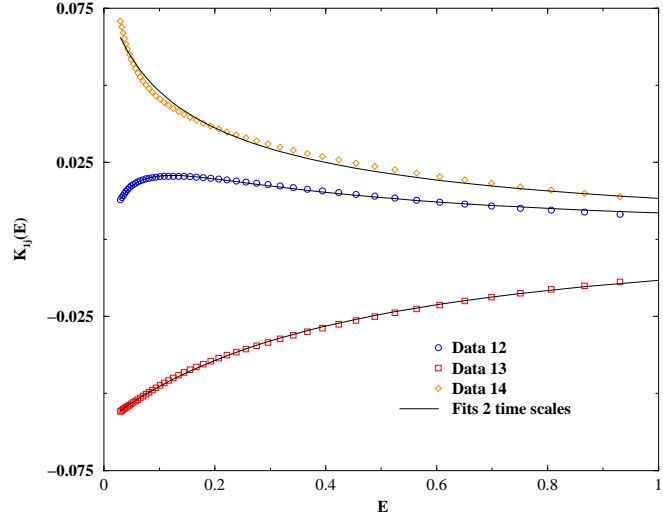,width=7cm,angle=270} 
\end{center}
\caption{Typical cross-influence kernels $K_{ij}(E)$ for three pairs of
stocks, and 
fit with the Laplace transforms of the sum of two exponentials. Note that
the influence amplitudes have different signs, even for the same pair of
stock, depending on the time scale.
}
\label{Fig6}
\end{figure}


Coming back to the observation that the correlation matrix is frequency dependent, one should note that the
scale dependent correlation matrix $C_{ij}(\tau)$ is related to ${\cal C}_{ij}(\omega)$ by:
\be
C_{ij}(\tau)= \langle r_{i,\tau} r_{j,\tau} \rangle_c = \int \ d\omega 
\ S^2(\omega \tau) {\cal C}_{ij}(\omega)
\ee
where $S(.)$ is the form factor (i.e. Fourier transform of the window used to define returns on scale $\tau$, for
example a flat window in the simplest case). Therefore, for $\tau$ small one finds that residuals are uncorrelated
(i.e. the correlation matrix has no structure beyond the market mode):
\be
C_{ij}(\tau \to 0) \approx D_i \delta_{ij},
\ee
whereas on long time scales the full correlation develops as:
\be
C_{ij}(\tau \to \infty) \approx D_i \delta_{ij} + \int_0^\infty {\rm d}\tau \left[D_j K_{ij}(\tau)+D_i K_{ji}(\tau)\right].
\ee
The emergence of correlation structure therefore reveals the lagged cross-influences in the market. Note that on
long time scales, small $K^0$'s can be counterbalanced by large synchronisation times $\Gamma^{-1}$, and lead to 
significant correlations between `weakly coupled' stocks.

We believe that a more systematic empirical study of the influence matrix $K_{ij}$ and the way it should be cleaned,
in the spirit of the discussion in section II, is worth investigating in details. 

\vskip 0.3cm
\small{
We want to thank Pedro Fonseca and Boris Schlittgen for many discussions on the
issues addressed in sections IV and V, and Szilard Pafka and Imre Kondor for sharing the
results on the EWMA matrices given in section III. We also thank G\'erard 
Ben Arous and Jack Silverstein for several clarifying discussions. We also thank the 
organisers of the meeting in Cracow for inviting us and for making 
the conference a success. 
}

\end{document}